\documentclass[prl,twocolumn,showpacs,preprintnumbers,amsmath,amssymb,superscriptaddress,amsmath]{revtex4-1}
\makeatletter
\renewcommand*\env@matrix[1][c]{\hskip -\arraycolsep
  \let\@ifnextchar\new@ifnextchar
  \array{*\c@MaxMatrixCols #1}}
\makeatother

\usepackage{amsmath,array}
\usepackage{graphicx,subfigure}  
\usepackage{dcolumn}                 
\usepackage{bm}                          
\usepackage{multirow}

\graphicspath{{./}}

\begin{document}
  
\title{Comment on Phys. Rev. Lett. {\bf 108}, 191802 (2012): "Observation of Reactor Electron Antineutrino Disappearance in the RENO Experiment"}

\author{Thierry Lasserre}
\email{Corresponding author: thierry.lasserre@cea.fr}
\affiliation{Commissariat \`a l'Energie Atomique et aux Energies Alternatives,\\
Centre de Saclay, IRFU, 91191 Gif-sur-Yvette, France}
\affiliation{Astroparticule et Cosmologie APC, 10 rue Alice Domon et
  L\'eonie Duquet, 75205 Paris cedex 13, France}

\author{Guillaume Mention}
\affiliation{Commissariat \`a l'Energie Atomique et aux Energies Alternatives,\\
Centre de Saclay, IRFU, 91191 Gif-sur-Yvette, France}

\author{Michel Cribier}
\affiliation{Commissariat \`a l'Energie Atomique et aux Energies Alternatives,\\
Centre de Saclay, IRFU, 91191 Gif-sur-Yvette, France}
\affiliation{Astroparticule et Cosmologie APC, 10 rue Alice Domon et
  L\'eonie Duquet, 75205 Paris cedex 13, France}

\author{Antoine Collin}
\affiliation{Commissariat \`a l'Energie Atomique et aux Energies Alternatives,\\
Centre de Saclay, IRFU, 91191 Gif-sur-Yvette, France}

\author{Vincent Durand}
\affiliation{Commissariat \`a l'Energie Atomique et aux Energies Alternatives,\\
Centre de Saclay, IRFU, 91191 Gif-sur-Yvette, France}
\affiliation{Astroparticule et Cosmologie APC, 10 rue Alice Domon et
  L\'eonie Duquet, 75205 Paris cedex 13, France}

\author{Vincent Fischer}
\affiliation{Commissariat \`a l'Energie Atomique et aux Energies Alternatives,\\
Centre de Saclay, IRFU, 91191 Gif-sur-Yvette, France}

\author{Jonathan Gaffiot}
\affiliation{Commissariat \`a l'Energie Atomique et aux Energies Alternatives,\\
Centre de Saclay, IRFU, 91191 Gif-sur-Yvette, France}

\author{David Lhuillier}
\affiliation{Commissariat \`a l'Energie Atomique et aux Energies Alternatives,\\
Centre de Saclay, IRFU, 91191 Gif-sur-Yvette, France}

\author{Alain Letourneau}
\affiliation{Commissariat \`a l'Energie Atomique et aux Energies Alternatives,\\
Centre de Saclay, IRFU, 91191 Gif-sur-Yvette, France}

\author{Matthieu Vivier}
\affiliation{Commissariat \`a l'Energie Atomique et aux Energies Alternatives,\\
Centre de Saclay, IRFU, 91191 Gif-sur-Yvette, France}

\date{\today}

\begin{abstract}
The RENO experiment recently reported the disappearance of reactor electron antineutrinos ($\bar{\nu}_e$) consistent with neutrino oscillations, with a significance of 4.9 standard deviations. The published ratio of observed to expected number of antineutrinos in the far detector is R=$0.920 \pm 0.009({\rm stat.}) \pm 0.014({\rm syst.})$ and corresponds to $\sin^2 2 \theta_{13} = 0.113 \pm 0.013({\rm stat.}) \pm 0.019({\rm syst.})$, using a rate-only analysis. In this letter we reanalyze the data and we find a ratio R=$0.903 \pm 0.01({\rm stat.})$, leading to $\sin^2 2 \theta_{13} = 0.135$. Moreover we show that the $\sin^2 2 \theta_{13}$  measurement still depend of the prompt high energy bound beyond 4 MeV, contrarily to the expectation based on neutrino oscillation.
\end{abstract}

\maketitle

\subsection{Introduction}
\label{intro}

\noindent For $\sim$1-2 km baseline reactor neutrino experiments the survival probability of electron antineutrinos ($\bar{\nu}_e$) is 
\begin{equation}
 P_{survival} \approx 1- \sin^2 2 \theta_{13} \sin^2(1.27 \Delta m^2 L/E),           
\end{equation}
where $E$ is the energy of $\bar{\nu}_e$'s in MeV, and $L$ is the baseline distance in meters between the reactor and detector and $\Delta m^2 = (2.32_{-0.08}^{+0.12})\times 10^{-3}$ eV$^2$ \cite{minosdm2}. \\

\noindent In 2011, first indications of a non-zero $\theta_{13}$ value have been reported by the T2K \cite{t2k} and MINOS \cite{minos} accelerator appearance experiments, and by the Double Chooz reactor disappearance experiment \cite{dc}. Very recently, the Daya Bay experiment reported the most precise measurement of $\theta_{13}$ using a rate-only analysis and found $\sin^22\theta_{13}=0.092\pm 0.016({\rm stat})\pm0.005({\rm syst})$ \cite{dayabay}. 

\subsection{RENO Oscillation Results}
\label{reno}

\noindent In a first publication released on April $3^{rd}$ 2012 the RENO collaboration reported a measurement of the neutrino oscillation mixing angle $\theta_{13}$, based on the observed $\bar{\nu}_e$ disappearance with a significance of 6.3 standard deviations \cite{renov1}. 
\noindent They obtained a ratio of observed to expected number of $\bar{\nu}_e$ in the far detector of $R=0.922 \pm 0.010({\rm stat.}) \pm 0.008({\rm syst.})$. Using a rate-only analysis, they derived a very precise measurement of $\sin^2 2 \theta_{13} = 0.103 \pm 0.013({\rm stat.}) \pm 0.011({\rm syst.})$.\\

\begin{figure}[!ht]
\begin{center}
\includegraphics[scale=0.42]{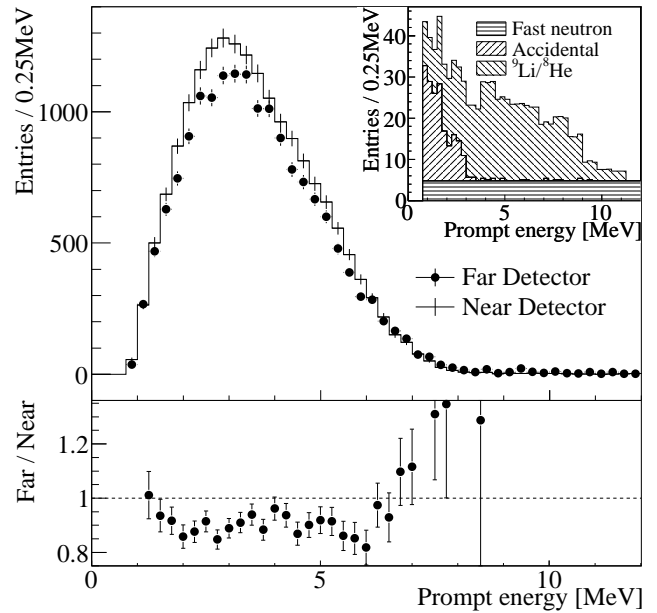}
\caption{\label{f:fig1}
Original figure taken from \cite{renov2}. Upper panel: spectrum of the prompt signal observed in the far detector compared with the non-oscillation prediction from the $\bar{\nu}_e$ spectrum observed in the near detector. The backgrounds shown in the inset are those subtracted to the far spectrum. Errors are statistical uncertainties only. Lower panel: ratio of the two spectra displayed above.}
\end{center}
\end{figure}

\noindent On April $8^{th}$ 2012, in a second version of their publication, the RENO collaboration revised their results \cite{renov2}. The ratio of observed to expected number of $\bar{\nu}_e$'s in the far detector has been updated to R=$0.920 \pm 0.009({\rm stat.}) \pm 0.014({\rm syst.})$, though the $E_{\rm prompt}$ energy range being used for the computation is not explicitely mentioned. This revision implies a modification of the measurement of $\theta_{13}$ leading to $\sin^2 2 \theta_{13} = 0.113 \pm 0.013({\rm stat.}) \pm 0.019({\rm syst.})$. It is worth noting that estimate of the total systematic uncertainty increased by a factor of 1.7 between the two release versions. The collaboration updated the results by explicitly including the global normalization in the oscillation fit (constrained by the absolute reactor neutrino flux uncertainty of 2.5\%) and modifying the lithium background determination. The detail characteristics of the RENO experiment can be found in \cite{renov1,renov2}.
     
\subsection{Expected disappearance signal}
\label{reana}

\noindent In order to assess the expected disappearance signal at the far detector we perform a simulation of the RENO experiment. We use the fluxes published in \cite{renov1} to qualify our simulation setup with six nuclear cores and two detectors, within the percent level.
\noindent Table \ref{tab:signal} shows the RENO expected cumulative $\bar{\nu}_e$ deficit, $\delta(E_{\rm high})$, as a function of the high prompt energy bound, $E_{\rm high}$. It is given by the formula
\begin{equation}
\delta(E_{\rm high}) = \frac{\int_{1 \rm MeV}^{E_{\rm high}}S(E)(1-P_{survival}(E_\nu(E)))dE}{\int_{1 \rm MeV}^{E_{\rm high}}S(E)dE}, 
\label{signaldef}
\end{equation}
where S(E) is the expected prompt energy spectrum at the far detector, taking into account the the six reactor distances, powers and fuel compositions. 
It shows that roughly 98\% of the $\bar{\nu}_e$ deficit ('the oscillation signal') is expected below 6 MeV (prompt energy). The oscillation analysis results must be thus stable by changing the high energy bound to values between 6 and 12 MeV.
\begin{table}[hbt]
 \begin{center}
 \begin{tabular*}{0.48\textwidth}{@{\extracolsep{\fill}} l c c c c c}
 
 \hline
      $E_{\rm high}$ (MeV)                    & 3 & 4 & 5 & 6 & 8   \\
 \hline
     Cumulative deficit fraction (\%)      & 49.5 & 78.2 & 92.6 & 98.1 & 99.97  \\    
 \hline 
 \end{tabular*}
 \end{center}
  \caption{Simulation of the RENO cumulative $\bar{\nu}_e$ deficit in the prompt energy range [1;$E_{\rm high}$] MeV, assuming $\Delta m^2 = 2.32\times 10^{-3}$ eV$^2$ .  Most of the oscillation signal is expected below 5 MeV. }
 \label{tab:signal}
 \end{table}
\subsection{Re-analysis of RENO data}
\noindent In this section, we present a reanalysis of the RENO published data. We use the Far/Near ratio spectrum (bottom frame of Figure \ref{f:fig1}).  We recompute the mean Far/Near ratio over the $E_{prompt}$ = [1-6] MeV energy range, and unexpectedly obtain a lower Far/Near ratio R=$0.903 \pm 0.010({\rm stat.})$, 2\% downward shifted with respect to the RENO published version. 
We then evaluate the impact of this bias on the oscillation result.  We simulate the relation between the Far/Near ratio and $\sin^2 2 \theta_{13}$ for a rate-only analysis, leading to the expression:
\begin{equation}
\sin^22\theta_{13} \simeq \frac{1-<P_{survival}>}{<\sin^{2}(\Delta m^2L/4E)>} \sim \frac{1-R}{0.715}, 
\label{Rtheta13}
\end{equation}
valid for $E_{\rm high}>$5 MeV. 
\noindent Our reanalysis of the RENO data in the $E_{prompt}$ = [1-6] MeV energy range lead to $\sin^22\theta_{13}=0.135$. Consequently, the data indicates a possible underestimation of $\sin^22\theta_{13}$ by about 20\% (1$\sigma$). \\
\begin{figure}[!htb]
\begin{center}
\includegraphics[scale=0.45]{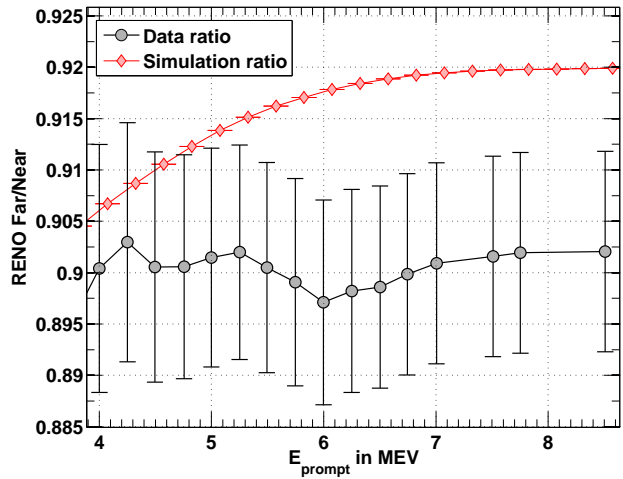}
\caption{\label{f:fig2} Weighted averaged of the RENO Far/Near data (bottom frame of Figure \ref{f:fig1}) as a function of the high energy bound (statistical errors only).  The red points shows the expected behavior based on our RENO simulation ($\sin^{2}(2\theta_{13})$=0.113), whereas the black points represent our RENO reanalysis. Our measured Far/Near ratio is 20\% lower than the RENO published value, R=0.920. Moreover, the data do not match well expected evolution for neutrino oscillation.}
\end{center}
\end{figure}
\\
\noindent We now test the consistency of the data points and their error bars (bottom frame of Figure \ref{f:fig1}). The evolution of the Far/Near ratio as a function of the high energy bound of the prompt energy range, $E_{high}$ is displayed in Figure \ref{f:fig2}. Consequently we study the evolution of the $\sin^{2}(2\theta_{13})$ measurements as a function of $E_{high}$. We retrieve the data from the upper panel of Figure \ref{f:fig1}, presenting the far detector spectrum and the expected near detector non-oscillating spectrum. The upper panel shows background subtracted spectra up to 12 MeV and then allows us building the Far/Near ratio as a function of the prompt energy up to 12 MeV. In the [1-8] MeV energy range we cross check our results using the top and bottom frames of Figure \ref{f:fig1}, validating the method of properly normalizing both near and far spectra based on published distances, live times and efficiencies \cite{renov1,renov2}. Results are displayed on Figure \ref{f:fig3} where the simulation is displayed in red and the data reanalysis in black. The evolution on $\sin^{2}(2\theta_{13})$, does not match the expectation were 80\% of the entire $\bar{\nu}_e$ deficit is expected below 4 MeV. This implies that the measurement of $\sin^{2}(2\theta_{13})$ should not significantly change with an higher energy bound contrary to the pathologic behavior of the data leading for instance to a modification of $\sin^{2}(2\theta_{13})$ value from 0.11 to 0.135 between 4 and 6 MeV, whereas the $\bar{\nu}_e$ expected deficit varies by less than 2\%.\\
\begin{figure}[!h]
\begin{center}
\includegraphics[scale=0.45]{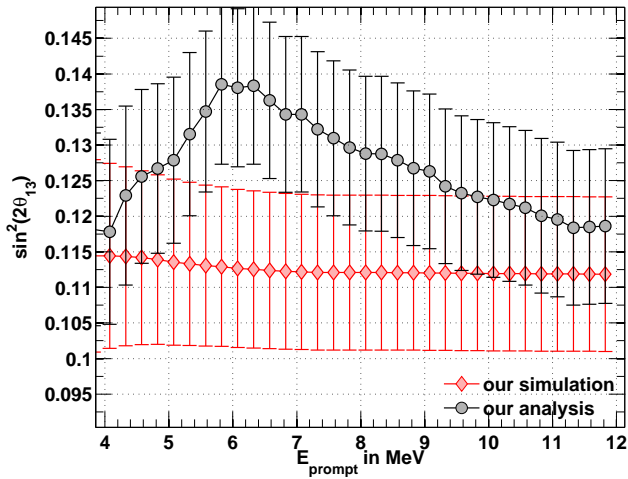}
\caption{\label{f:fig3} Simulation and reanalysis of the $\sin^22\theta_{13}$ constraint as a function of the high energy bound of the prompt energy range over which the Far/Near ration is computed (for $\sin^{2}(2\theta_{13})$=0.113, statistical errors only). We use the data published in the upper panel of Figure \ref{f:fig1}. Accounting 99\% of the neutrino oscillation signal, at $E_{high}$=6 MeV, one obtains $\sin^22\theta_{13}=0.135$.  But the measured central value $\sin^22\theta_{13}$ still decreases towards 0.115 if we (erroneously) include data over the whole prompt energy range, [1-12] MeV, contrarily to neutrino oscillation expected behavior.}
\end{center}
\end{figure}

\noindent Surprisingly the rate-only measurement of $\sin^22\theta_{13}$ still changes when considering energy bound higher than 8 MeV, as opposed to expectations (red curve on Figure \ref{f:fig3}). This change is clearly not due to neutrino oscillation physics. Using $E_{high}$=12 MeV one obtains $\sin^22\theta_{13}=0.115$, consistent with the value published by the RENO collaboration \cite{renov2}. We conclude that the deficit above 6 MeV could be interpreted as Far/Near detector response differences or as an inaccurate background subtraction.\\

\noindent As a cross-check of the method developed to reanalyze the recent RENO reactor antineutrino results \cite{renov2}, we recompute the Daya Bay Far/Near ratio based on published information appearing in the Figure 5 (bottom frame) of \cite{dayabay}. The detail characteristics of the experiment can be found in \cite{dayabay}. We find an averaged Daya Bay Far/Near ratio of $R=0.937\pm 0.011({\rm stat})$ in the $E_{prompt}$ = [1-7.8] MeV energy range, which is in excellent agreement with the Daya Bay published result. Furthermore, we study the Far/Near ratio as a function of the high energy bound of the prompt energy range, $E_{high}$. The Far/Near ratio is stable when accounting data with $E_{high}>$5 MeV, consistent with neutrino oscillations.

\subsection{Conclusion}
\label{conclusion}

\noindent In this letter, we reanalyze the reactor neutrino data recently published by the RENO collaboration \cite{renov2}. While the collaboration report a deficit of 8.0\% in the far detector compared to what is expected in the near detector if no-oscillation, our analysis, which considers the expected neutrino oscillation energy range ($E_{prompt}$ = [1-8] MeV), leads to a deficit of 10.0\%.  Our reanalysis points out to a possible bias of the central value by about +20\% (1$\sigma$), leading to a higher $\sin^2 2 \theta_{13} = 0.135$. This new best fit result is more than 2$\sigma$ off the Daya Bay central value. We then study both the Far/Near ratio and the $\sin^22\theta_{13}$ measurements as a function of the high energy bound of the prompt energy range, $E_{high}$. The data shows a pathologic behavior, especially beyond 5 MeV, which could be interpreted as an underestimation of the relative systematic uncertainty between the near and far detectors or as an inaccurate background subtraction. Finally, we attempt to understand the origin of this $\sin^2 2 \theta_{13}$ best fit discrepancy and we notice that a fallacious inclusion of the data between 8 and 12 MeV in the oscillation analysis (though no oscillation should occur in this energy range) would lead to $\sin^2 2 \theta_{13} = 0.115$. We kindly ask the collaboration to shed more light on their analysis in a detailed publication.

\subsection{Acknowledgements}
\noindent We thank H. de Kerret C. Mariani and S. Schoenert for useful discussions.


\end{document}